\documentclass{article}
\title{Using Compton Imaging to Locate Moving Gamma-Ray Sources in the GRETINA Detector}
\author{Dr. Robert Crabbs \\ University of California, Berkeley \\
	\and 
	Dr. I-Yang Lee \\ Lawrence-Berkeley National Laboratory \\
	\and 
	Dr. Kai Vetter \\	Lawrence-Berkeley National Laboratory \\
	}
\date{\today}

\usepackage{amsmath} 
\usepackage{bm} 
\usepackage{graphicx} 
\usepackage{pdfpages} 
\usepackage{textcomp} 
\usepackage{color} 
\usepackage{multicol,caption} 
\usepackage{geometry} 
\usepackage{hyperref} 

\hypersetup{
  colorlinks   = true, 
  urlcolor     = blue, 
  linkcolor    = black, 
  citecolor   = red 
}
\newcommand{\norm}[1]{\lVert#1\rVert}

\newenvironment{multicolfigure}
  {\par\medskip\noindent\minipage{\linewidth}}
  {\endminipage\par\medskip}

\begin{document}

\maketitle

\begin{abstract}
GRETA, the \textbf{G}amma-\textbf{R}ay \textbf{E}nergy \textbf{T}racking \textbf{A}rray, is an array of highly-segmented HPGe detectors designed to track $ \gamma $-rays emitted in beam-physics experiments. Its high detection efficiency and state-of-the-art position resolution make it well-suited for Compton imaging applications. In this paper, we use simulated imaging data to illustrate how GRETA can be used to locate the emission points of photons produced during beam experiments. This lays the groundwork for nuclear lifetime measurements using Compton imaging. 
\end{abstract}

\begin{multicols}{2}


\section{Introduction}
\label{sec:introduction}

Gamma-ray tracking \cite{gamma_ray_tracking_detectors} \cite{new_concepts_in_gamma_detection} is a major advance in gamma-ray spectroscopy. A 4$ \pi $ tracking-array would be a powerful instrument for a broad range of experiments in low-energy nuclear science \cite{nsac_long_range_plan} \cite{nupecc_long_range_plan}, especially for the nuclei far from the line of stability. Developments of these instruments are underway \cite{agata_and_greta} both in the US (GRETINA/GRETA) \cite{gretina_performance_whitepaper} \cite{gretina_spectroscopy_performance} \cite{greta_website} and Europe (AGATA) \cite{agata_whitepaper} \cite{agata_website}. The GRETA collaboration has built a partial array, called the Gamma-Ray Energy Tracking IN-beam Array (GRETINA). The array's primary purpose is for high-efficiency, high-precision $ \gamma $-ray spectroscopy. Its position sensitivity allows the sequencing of photon tracks via Compton kinematics, which is used, among other things, to estimate the emission angle of a gamma ray relative to its parent nucleus’s velocity. This is key for spectroscopy because it allows us to correct Doppler shifts in the lab-frame photon energy.
\par
GRETINA's excellent position- and energy-resolution also open the door for gamma-ray imaging. \cite{gamma_ray_tracking_opportunities} If we can accurately reconstruct the photon source distributions from beam experiments, we can measure nuclear lifetimes more efficiently than is possible with current methods such as RDM. \cite{lifetime_measurement_74rb} \cite{rdm_triplex_plunger} Where RDM generally requires multiple measurements to find a lifetime, imaging can theoretically deliver full de-excitation curves from a single run. Lifetimes can be extracted directly by fitting to these curves.
\par
Our goal here is to understand the imaging performance that can be obtained using Compton imaging in GRETINA. Imaging resolution and efficiency will ultimately define the conditions for which we can measure lifetimes with this technique.


\section{Compton Cones}
\label{sec:compton_cones}

Our goal with Compton Imaging is to locate the emission points of $ \gamma $-rays created during in-beam experiments. Each detected photon leaves a track of interactions in the detector. Using Compton kinematics and the first \& second interaction points in the track, we can define a ``cone'' of possible directions from which that photon originated. From there, we can calculate the photon's emission point $ \bm{X_0} $ as one of the intersections of this cone with the beamline. (See Figure \ref{fig:compton_imaging_with_errors}A)
\par
A Compton cone is defined by its vertex $ \bm{X_C} $, its central axis $ \bm{V_C} $, and its opening angle $ \theta_C $. These parameters can be computed from a photon track once we know its correct sequence. \cite{compton_sequencing_in_gretina} \cite{intersection_line_cone} Let this sequence of photon interactions have positions $ \bm{X_1}, ..., \bm{X_N} $ in the detector, with corresponding energy depositions $ {\Delta}E_1, ..., {\Delta}E_N $. 
\par
To get the Compton cone's opening angle, consider the total energy of the photon track:
\begin{equation}
E_{total} = \sum_{n=1}^{N} \Delta E_n
\end{equation}
This is the initial energy of the photon before it enters the detector, assuming the photon deposits its full energy. Because we also know the energy deposition at the first scattering point $ \bm{X_1} $, we can calculate $ \mu_1 $, the cosine of the initial scattering angle, from Compton kinematics:
\begin{align}
\label{eq:mu_calc}
\mu &= 1 - {m_e}c ^2 \left( \frac{1}{E_f} - \frac{1}{E_i} \right) \\
&= 1 - \frac{{m_e}c^2{\Delta}E}{E_i (E_i - {\Delta}E)} \\
\mu_1 &= 1 - \frac{{m_e}c^2{\Delta}E_1}{E_{total} (E_{total} - {\Delta}E_1)}
\end{align}
This gives the cosine of the angle between the photon's final and initial headings at $ \bm{X_1} $. Note that the final heading is simply defined as a unit vector from $ \bm{X_1} $ towards $ \bm{X_2} $, i.e.:
\begin{equation}
\bm{\hat{V}_1} = \frac{\bm{X_2} - \bm{X_1}}{\norm{\bm{X_2} - \bm{X_1}}}
\end{equation}
We can thus say that:
\begin{equation}
\label{eq:initial_photon_header}
\bm{\hat{V}_0} \cdot \bm{\hat{V}_1} = \mu_1
\end{equation}
where $ \bm{\hat{V}_0} $ is the unit vector that defines the photon's incident heading on $ \bm{X_1} $.
\par
Equation \ref{eq:initial_photon_header} defines the \textit{cone} of possible directions for the incident photon vector. The cone's opening angle is $ \theta_C = \cos^{-1}{\mu_C} = \cos^{-1}{\mu_1} $, while its vertex is simply $ \bm{X_C} = \bm{X_1} $. The cone's axis is the unit vector from the \textit{second} interaction to the first -- in other words, $ \bm{\hat{V}_C} = -\bm{\hat{V}_1} $. Note that the photon must have originated from a point somewhere on the \textit{forward} sheet of this cone.


\section{Locating Photons Emitted Along a Beam}
\label{sec:beam_cone_intersections}

The next step is determining where each Compton cone intersects the GRETINA beamline. In other Compton imaging applications, the distance from the detector to the emission point is often unknown, limiting results to an angular distribution. (See References \cite{filtered_backprojection}, \cite{compton_electrons_and_filtered_backprojection}, and \cite{compton_imaging_in_agata} for details on stereographic projections and Compton imaging.) However, GRETINA's source geometry is uniquely constrained.
\par
In lifetime experiments, an accelerator delivers high-energy projectiles to a thin target, creating excited nuclei through fusion or other nuclear reactions. These products recoil out of the target in a highly-collimated beam traveling at relativistic speeds (typically 0.05-0.4c, depending on the beam energy and the kinematics of the nuclear reaction). The recoil nuclei de-excite a short time later, emitting one or more characteristic gamma-rays. This means, for our imaging problem, we can assume our photons originate somewhere along the beamline, downstream of the target. This is much easier than trying to locate an unconstrained point source in 3D.
\par
\begin{figure}[ht]
\centering
\includegraphics[width=0.75\textwidth]{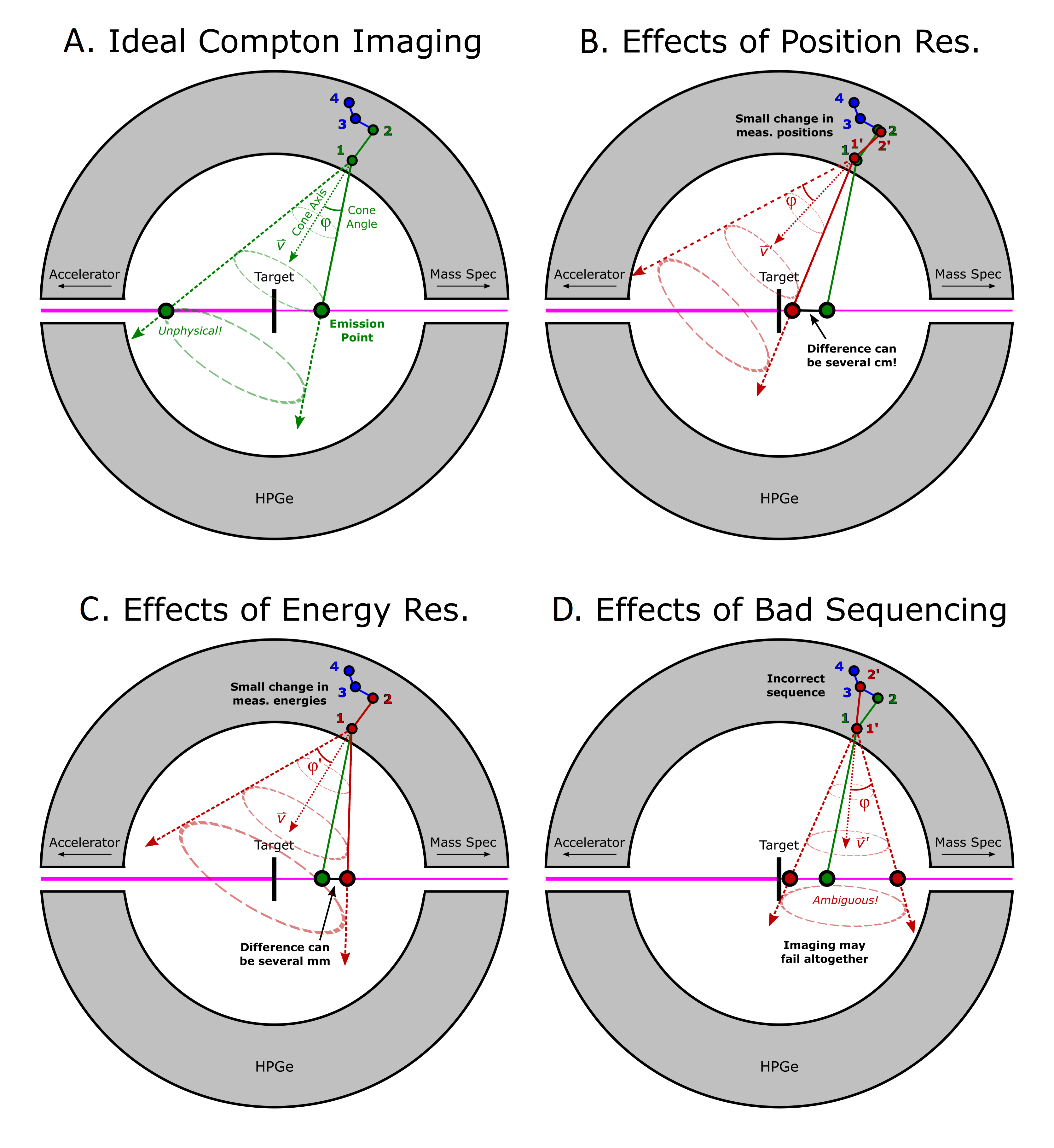}
\caption{Compton imaging under ideal \& non-ideal detector resolution}
\label{fig:compton_imaging_with_errors}
\end{figure}

As discussed above, Compton imaging yields a cone of possible directions from which a detected photon may have originated. Assuming the parent nuclei form a pencil beam, each Compton cone can intersect the beam at a maximum of two locations. Ideally, one of these intersections will be \textit{upstream} of the target, which we can rule out as physically impossible. This will yield a single unambiguous photon origin -- the intersection downstream of the target. In many cases, however, both intersections are downstream of the target. For our research, we've opted to throw out such ambiguous data points. Figure \ref{fig:compton_imaging_with_errors}A shows the ideal Compton cone geometry in GRETA, the full $ 4 \pi $-version of GRETINA.
\par
The detailed derivation of the cone-beam intersection points can be found in Reference \cite{cone_beam_intersections_in_gretina}. In the end, we arrive at a quadratic equation that depends on the beam axis and the Compton cone axis, vertex, and opening angle. The intersections are $ \bm{X_{0,1}} = t_1 \bm{\hat{B}} $ and $ \bm{X_{0,2}} = t_2 \bm{\hat{B}} $, where $ t_1 $ and $ t_2 $ are the solutions to that equation and $ \bm{\hat{B}} $ is the beam axis.
\par
While on paper this is very promising, Compton imaging with real detectors is not nearly so clean. Imperfect detector position and energy resolution introduces uncertainty around the interaction coordinates $ \bm{X_1} $, ..., $ \bm{X_N} $ and energy depositions $ E_1 $, ..., $ E_N $. \cite{position_sensitivity_in_hpge} \cite{greta_detector_performance} Both the Compton cone axis ($ \bm{V_C} = \bm{X_1} - \bm{X_2} $) and the cone angle (Equation \ref{eq:mu_calc}) are directly affected by these quantities. Figures \ref{fig:compton_imaging_with_errors}B-D illustrate the effects that imperfect detector performance can have on the imaging problem. 
\par
Energy resolution changes the cone angle, but the difference is generally small. Consider a 1.0 MeV photon that initially scatters at 45\textdegree{} in the detector, leaving a 364 keV energy deposition at $ \bm{X_1} $. Suppose with 0.2\% energy resolution, this deposition registers as 363 keV while the full photon energy registers 1002 keV. The scattering angle computed in this scenario would then be 44.7\textdegree{}, shifting the cone-beam intersection by approximately 0.3\textdegree{}. Depending on the orientation and location of the cone relative to the beam axis, this angular shift could correspond to several millimeters or more.
\par
Position resolution creates much more serious problems for Compton imaging. Figure \ref{fig:approx_angular_error_compton} shows a simple estimate for angular resolution based on errors in the Compton cone axis. Consider a track where the first two hits are $ \bm{X_1} $ and $ \bm{X_2} $. Suppose the detector registers hits at $ \bm{X_1'} $ and $ \bm{X_2'} $, which are a distance $ L' $ apart. (Here, $ L' $ is referred to as the ``Compton lever arm''.) In the worst case, the shifts are perpendicular to the \textit{old} cone axis $ \bm{V_C'} = \bm{X_1'} - \bm{X_2'} $. The maximum angular error can be expressed as:
\begin{equation}
\label{eq:approx_angular_error_compton}
\Delta\theta_C = \arctan{\sqrt{2}\sigma_{xyz}/L'}
\end{equation}
where $ \sigma_{xyz} $ is the position resolution and $ L' = \norm{\bm{X_1} - \bm{X_2}} $. With typical $ \sigma_{xyz} = $ 3.0 mm position resolution and a measured lever arm of $ L' = $ 30.0 mm, the shift is 8.0\textdegree{}. At a 90\textdegree{} nominal emission angle and a cone vertex 180 mm from the beamline, this corresponds to a 25.2 mm shift from the true emission point.

\begin{multicolfigure}
\centering
\includegraphics[width=0.8\textwidth]{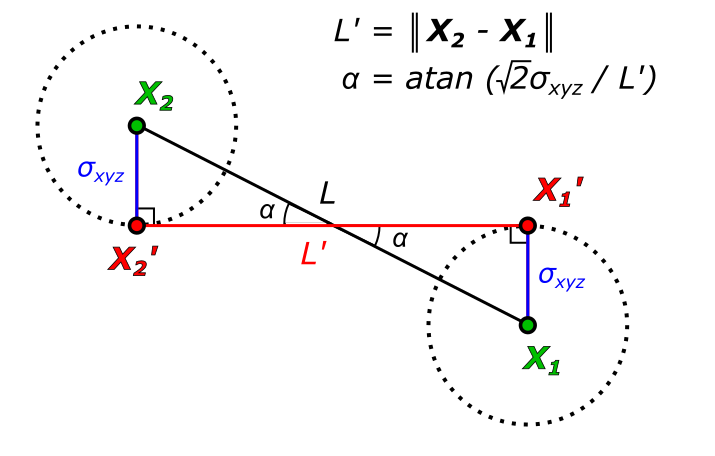}
\captionof{figure}{Analytic estimates of imaging resolution}
\label{fig:approx_angular_error_compton}
\end{multicolfigure}

\par
Inaccurately identifying either $ \bm{X_1} $ or $ \bm{X_0} $ in the interaction sequence will also shift the cone axis. These errors can be quite severe, as the distance between points is generally several times greater than the position resolution.
\par
Real-world beam properties are another source of error for our imaged source distributions. Our analysis here assumes that nuclei travel directly along the $ z $-axis, but real ion beams have a finite angular divergence and spot size on target. Typically, the divergence is $ < $ 1\textdegree{} and the spot size can be modeled as a 2D Gaussian with $ \sigma \leq $ 4 mm in both dimensions. With the proper instrumentation downstream of GRETINA, it is possible to measure the true, event-by-event headings of the recoil nuclei and correct for these effects. \cite{microchannel_plate_detectors} 
\par
In all these cases, it is possible for a Compton reconstruction to yield ambiguous, incorrect, or unphysical solutions, or no solutions at all. (A Compton cone may not intersect the beam-line when it is too far off-axis.) The imaging errors become more pronounced at shallower emission angles. As the distance between the cone vertex and the calculated emission point increases, the angular resolution translates to larger shifts along the beam-line.

\section{Detector Performance and Imaging Resolution}
\label{sec:compton_detector_performance}

Detector resolution, beam velocity, and the photon energy in the CM-frame are all factors that can affect Compton imaging resolution. To quantify their effects, we used a detailed Geant4 model to generate simulated photon interactions within the GRETINA array. \cite{UCGretina_model} This allowed us to incorporate realistic detector geometry, materials, Doppler shift, and beam divergence into our study of imaging resolution. The simulated photons were emitted by a beam of hypothetical recoil nuclei traveling at 0.3 \textpm{} 0.001c along the detector's $ z $ -axis. The recoil beam was given a divergence of $ \sigma_{\theta} = $ 0.25\textdegree{} and spot size of $ \sigma_{beam,x} = \sigma_{beam,y} = $ 0.1 mm at the target, which was placed at z = 0.0 mm. While we used multiple simulated lifetimes in our study, we focused on 2.7525 ns because that gave the recoil nuclei a space of 3 half-lives before they exited the central detector cavity at $ z = $ 180.0 mm. The photons were emitted with 1.0 MeV in the CM-frame of their parent nuclei. 
\par
The Geant4 model gives the true sequence, interaction locations, and energy depositions for each simulated photon track. It also provides the true emission point $ \bm{X_{0,true}} $ for each photon. The simulations are thus a good benchmark of our Compton imaging algorithm, which uses the principles in Section \ref{sec:beam_cone_intersections} to calculate where a photon originated from along the beam-line (denoted $ \bm{X_{0,calc}} $). With perfect simulated detector resolution (i.e. no uncertainty about the locations or energy of each interaction), we are able to correctly reconstruct most emission points. As noted above, sometimes both beam-cone intersections are downstream of the target, so even with perfect detector resolution we cannot get a definitive emission point for a photon. Reference \cite{compton_sequencing_in_gretina} provides more detail.  

\end{multicols}
\begin{multicols}{2}

\par
With the verified Compton tracking code, we then sought to quantify the effects of detector position and energy resolution on imaging resolution. Our first approach was based on error propagation for the cone-beam intersections, governed by the equations derived in Reference \cite{cone_beam_intersections_in_gretina}. This gave us analytic estimates of imaging resolution of the form:
\begin{equation}
\sigma_{img}^2 = \sigma_{img,xyz}^2 \sigma_{xyz}^2 + \sigma_{img,E}^2 \sigma_{E}^2
\end{equation}
where $ \sigma_{img} $ is the total effective imaging resolution, $ \sigma_{img,xyz} $ is the contribution due solely to detector position resolution $ \sigma_{xyz} $, and $ \sigma_{img,E} $ is the contribution due solely to detector energy resolution $ \sigma_{E} $. Figure \ref{fig:analytic_errors_colormap_compton} shows the normalized probability distribution of these analytic errors for a set of simulated photon tracks, using a detector of $ \sigma_{xyz} = $ 3.0 mm position- and $ \sigma_E = $ 2.0 keV energy-resolution. The colormap gives the ratio of $ \sigma_{img,xyz} $ to $ \sigma_{img,E} $. It appears that a 1.0 mm position resolution has a roughly 20x greater impact on Compton imaging resolution than does a 1.0 keV energy resolution. 
\par
The second approach was stochastic in nature. As discussed above, the Geant4 model gave us the exact locations and energy deposition for each photon interaction in the detector volume. We randomly shuffled the interaction sequence in each respective track, and added Gaussian noise to the positions and energy depositions of each interaction. This noise simulated the detector's position and energy resolution. Note that detector resolution was applied uniformly to all interactions, regardless of their energy deposition. Generally, though, GRETINA's detector resolution is energy-dependent. A higher-energy interaction (e.x. 1.0 MeV) can usually be located with better precision than a lower-energy one (e.x. 200 keV). Energy resolution typically scales with photon energy $ E $ as $ \sqrt{E^2 + C} $, where C is a constant. We ignored this effect for simplicity.
\par
With detector resolution accounted for, the shuffled photon tracks were then sent through our Compton sequencing and image reconstruction algorithm. This produced a Compton cone and a calculated emission point $ \bm{X_{0,calc}} $ for each track. As Geant4 provided the true emission point $ \bm{X_{0,true}} $ for each track, we could directly obtain the imaging error $ \bm{\Delta X_0} = \bm{X_{0,calc}} - \bm{X_{0,true}} $ for each reconstruction. For simplicity, we ignored beam divergence and spot size when calculating the intersections of the Compton cones with the beam. Only the z-components of the imaging errors were used to evaluate imaging resolution. 

\begin{multicolfigure}
\centering
\includegraphics[width=1.0\textwidth]{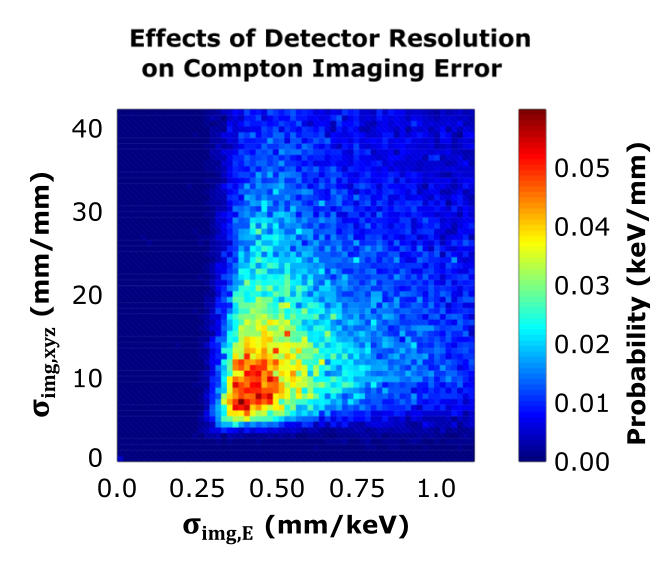}
\captionof{figure}{Analytic estimates of imaging resolution}
\textbf{Source}: $ E_{\gamma,CM} $ = 1000 keV, $ \beta $ = 0.3, $ \tau_0 $ = 3.0 ns \newline
\textbf{Resolution}: $ \sigma_{xyz} $ = 3.0 mm, $ \sigma_E $ = 2.0 keV
\label{fig:analytic_errors_colormap_compton}
\\~
\end{multicolfigure}

\par
The distribution of the imaging errors $ \bm{\Delta X_{0}} $ for all imaged photons gives the ``imaging response'' of the detector. Image resolution can be estimated by fitting Gaussians to the Compton imaging response, as illustrated in Figures \ref{fig:sample_compton_images_no_filters}A-C. Figure \ref{fig:sample_compton_images_no_filters}A shows the response for an ideal detector with perfect position \& energy resolution, limited only by Doppler Broadening effects; Figures \ref{fig:sample_compton_images_no_filters}B-C show responses for more realistic detectors with imperfect resolution. While these imaging responses are not completely Gaussian, such a simplification yields accurate FWHMs for the central peaks of the distributions. We use this to define the effective imaging resolution as $ \sigma_{img} = $ FWHM $ / 2 \sqrt{2 ln 2} $. Reference \cite{fitting_to_gaussians_in_matlab} provides details about the uncertainty calculations for the fits.
\par
In addition to resolution, imaging efficiency is another important factor for image quality. There are many reasons why a photon track might be rejected, thereby reducing the number of counts we have to construct the final Compton image. For example, reconstructions can result in ambiguous or unphysical emission points. A track may also be too short or too long for sequencing. Tracks with fewer than 3 interactions in the detector cannot be sequenced via Compton kinematics unless the emission point is already known (which defeats the purpose of imaging). On the other hand, the more interactions a track has, the more computationally-intensive it is to sequence. (This scales factorially.) For this study, we chose to accept tracks between 3 and 7 interactions in length.

\begin{multicolfigure}
\centering
\includegraphics[width=1.0\textwidth]{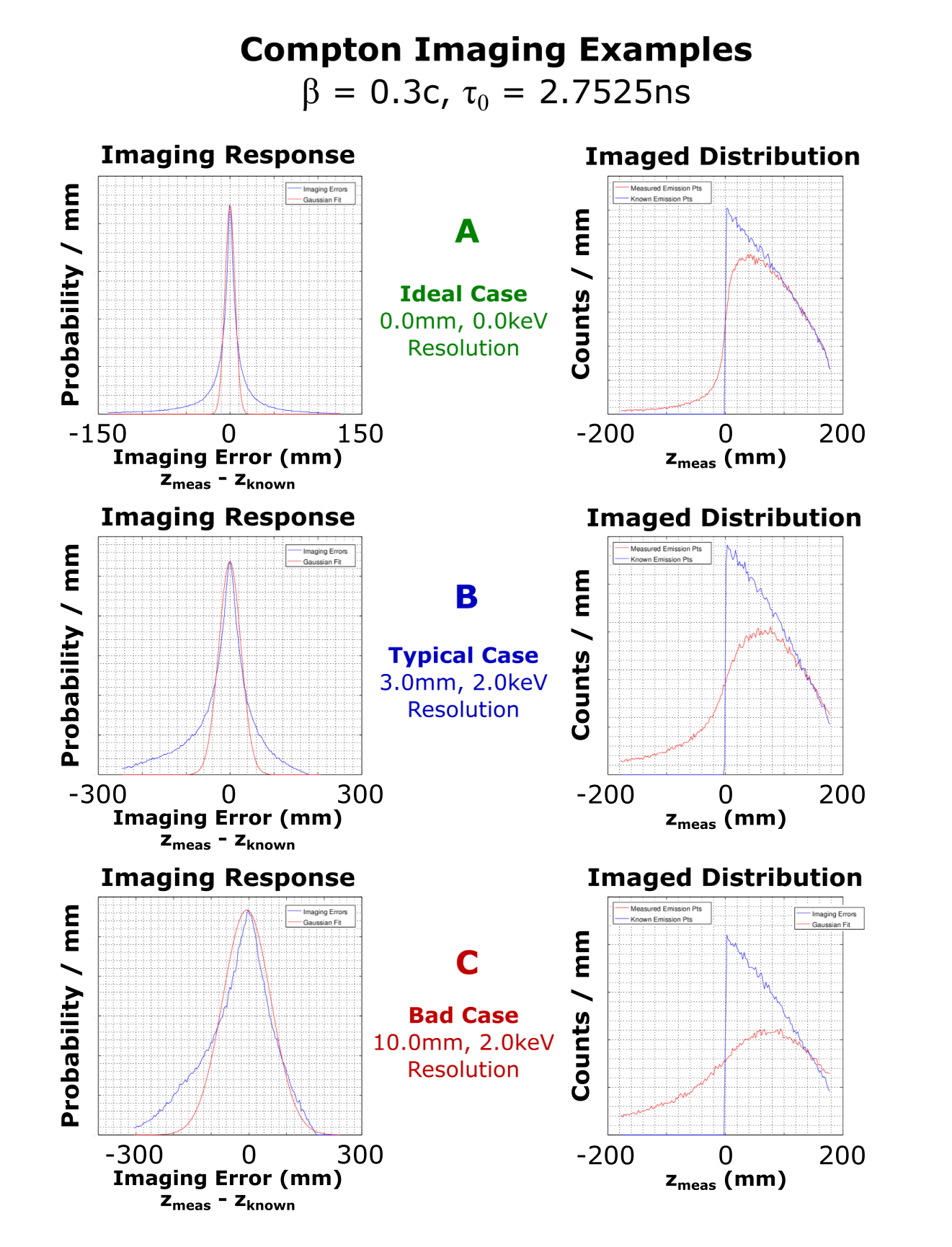}
\captionof{figure}{Sample Compton images}
For a range of detector position \& energy resolutions
\label{fig:sample_compton_images_no_filters}
\end{multicolfigure}

\par
We can also filter tracks by their measured lab-frame energy. This helps remove Compton background counts which deposited less than their full energy in the detector. There is a complication, however, because in lifetime measurements the photon sources are moving relativistically. For a recoil nucleus moving with speed $ \beta c $ in the lab frame, the emitted photon's energy can be calculated as:
\begin{equation}
E_{\gamma,Lab} = \frac{E_{\gamma,CM}}{\gamma (1 - \beta \cos{\phi})}
\end{equation}
where $ \gamma $ is the nucleus's Lorentz factor and $ \phi $ is the measured emission angle relative to the parent velocity in the lab frame. For characteristic $ \gamma $-rays in the MeV range, this means the photon's energy can be Doppler-shifted by up to several hundred keV in the lab frame. In order to identify partial energy depositions, then, we need to define a range of possible lab-frame energies based on the source velocity and known CM-frame energy of the photons. This range can be calculated from the minimum and maximum possible emission angles of the photon relative to the beam axis.
\par
Consider a track with its first interaction at $ \bm{X_1} $ in the detector. As mentioned previously, photons must be generated downstream of the beam target, \textit{typically} at (0 mm, 0 mm,  0 mm). For the sake of generality in our analysis, we chose to accept any photons emitted within GRETINA's central cavity, ranging from $ \bm{X_{0,L}} = $ (0 mm, 0 mm, -180 mm) to $ \bm{X_{0,R}} = $ (0 mm, 0 mm, 180 mm). The emission vectors $ \bm{V_{0,L}} = \bm{X_1} - \bm{X_{0,L}} $ and $ \bm{V_{0,R}} = \bm{X_1} - \bm{X_{0,R}} $ define the minimum and maximum possible emission angles, respectively, relative to the beam axis. This angular range corresponds to a range of possible lab-frame energies for a known characteristic photon energy. Any tracks outside that window can be rejected.
\par
Using the Geant4 model, we generated a total of 2.365M photon tracks with 1.0 MeV CM-frame energy. Of these, 1.529M tracks registered outside their possible Doppler-shifted energy ranges, meaning they deposited only partial energy in the detector. This left 0.836M plausible full-energy tracks for use in our image-quality tests. 
\par
Not all of these qualifying tracks make it into the final Compton images. As detector resolution degrades, for example, it becomes more likely for Compton imaging to produce ambiguous or unphysical emission points from a photon track. There are also certain data filters that will be described in Section \ref{sec:compton_imaging_filters} that can further reduce the number of counts in the output images. We define the imaging efficiency as the ratio $ \epsilon = N_{accepted} / N_{photopeak} $, where $ N_{accepted} $ represents the total number of qualifying tracks that returned unambiguous photon origins within GRETINA's central cavity, and $ N_{photopeak} $ is the total number of plausible full-energy counts registered by the detector.
\par
In general, position resolution dominates imaging performance for realistic values of energy resolution in HPGe. Tables 1-2 and Figures \ref{fig:compton_image_quality_vs_experimental_variables}D-E show the relative efficiencies as functions of detector position and energy resolution. (The numbers are normalized to the efficiency for a detector with perfect resolution). For example, at 3.0 mm position resolution, the output image has 3.8 times worse imaging resolution and contains only 70\% as many counts as the one produced by a detector with perfect position resolution. For Table 1, energy resolution was held constant at $ \sigma_E = $ 2.0 keV, while for Table 2, position resolution was held constant at $ \sigma_{xyz} = $ 3.0 mm. 
\par
While we did not study the effect in detail, the lab-frame energy of the photon may impact imaging quality. 200 keV photons yield fewer hits per track than 1.0 MeV ones, and the hits from the lower-energy tracks leave smaller energy depositions. This reduces not only the detector resolution but also the accuracy and efficiency of sequencing. Reference \cite{compton_sequencing_in_gretina} explores how photon energy affects sequencing and Compton imaging. 
\par
Detector geometry can also complicate the imaging problem. Consider a photon that is emitted near a detector (as opposed to one emitted near the center of the GRETINA inner cavity). The total distance traveled from emission to the first hit is then relatively small, which means angular resolution translates to a smaller spatial deviation on the beam-line. Nuclei with shorter lifetimes or lower recoil velocities will produces images closer to the beam target, and thus are expected to have somewhat worse imaging resolution.  
\par

\section{Imaging Filters}
\label{sec:compton_imaging_filters}

The factors discussed in Section \ref{sec:compton_detector_performance} are intrinsic to the experiment and the detector used. There are also multiple data filters that can be applied in post-processing to improve image quality. As with any data filter, though, there is a trade-off between data quality and imaging efficiency (i.e. the fraction of tracks used in the final image reconstruction). We studied the effects of filtering by Compton ``lever arm'' ($ L $), sequencing Figure-of-Merit ($ FoM $), and Doppler-corrected CM-frame energy ($ E_{CM} $). The sequencing figure-of-merit is defined:
\begin{equation}
\label{eq:sequencing_FoM_energy}
FoM = \frac{1}{(N_{hits} - 2)}{\sum_{j=2}^{N_{hits}-1} \frac{ ({\Delta}E_{meas,j} - {\Delta}E_{calc,j})^2}{E_{total}^2} }
\end{equation}
where $ E_{j,meas} $ and $ E_{j,calc} $ are the measured and calculated energies for the track's \textit{j}th photon hit, respectively, and $ E_{total} $ is the total measured track energy. \cite{compton_sequencing_in_gretina}
\par
We used the same simulation data as before, keeping detector position and energy resolution constant at $ \sigma_{xyz} = $ 3.0 mm and $ \sigma_E = $ 2.0 keV respectively. For each test, we modified one data filter at a time while keeping the others constant at their most ``permissive'' values: $ L \geq $ 0, $ FoM \geq $ 0, and $ E_{CM} \geq $ 0. We found that the Compton lever-arm filter has a much greater effect on image quality than does the sequencing Figure-of-Merit, but the Doppler-correction filter outperformed both by a wide margin. 
\par
As can be seen in Equation \ref{eq:approx_angular_error_compton}, the angular resolution in Compton imaging depends strongly on $ L' $, the distance between the first and second hits in a photon track (the Compton ``lever arm''). The larger $ L' $ is compared to $ \sigma_{xyz} $, the smaller the mean angular error is expected to be. Therefore, setting a minimum Compton lever arm is a convenient filter to remove tracks with high imaging uncertainty. 
\par
\begin{figure}[ht]
\centering
\includegraphics[width=1.0\textwidth]{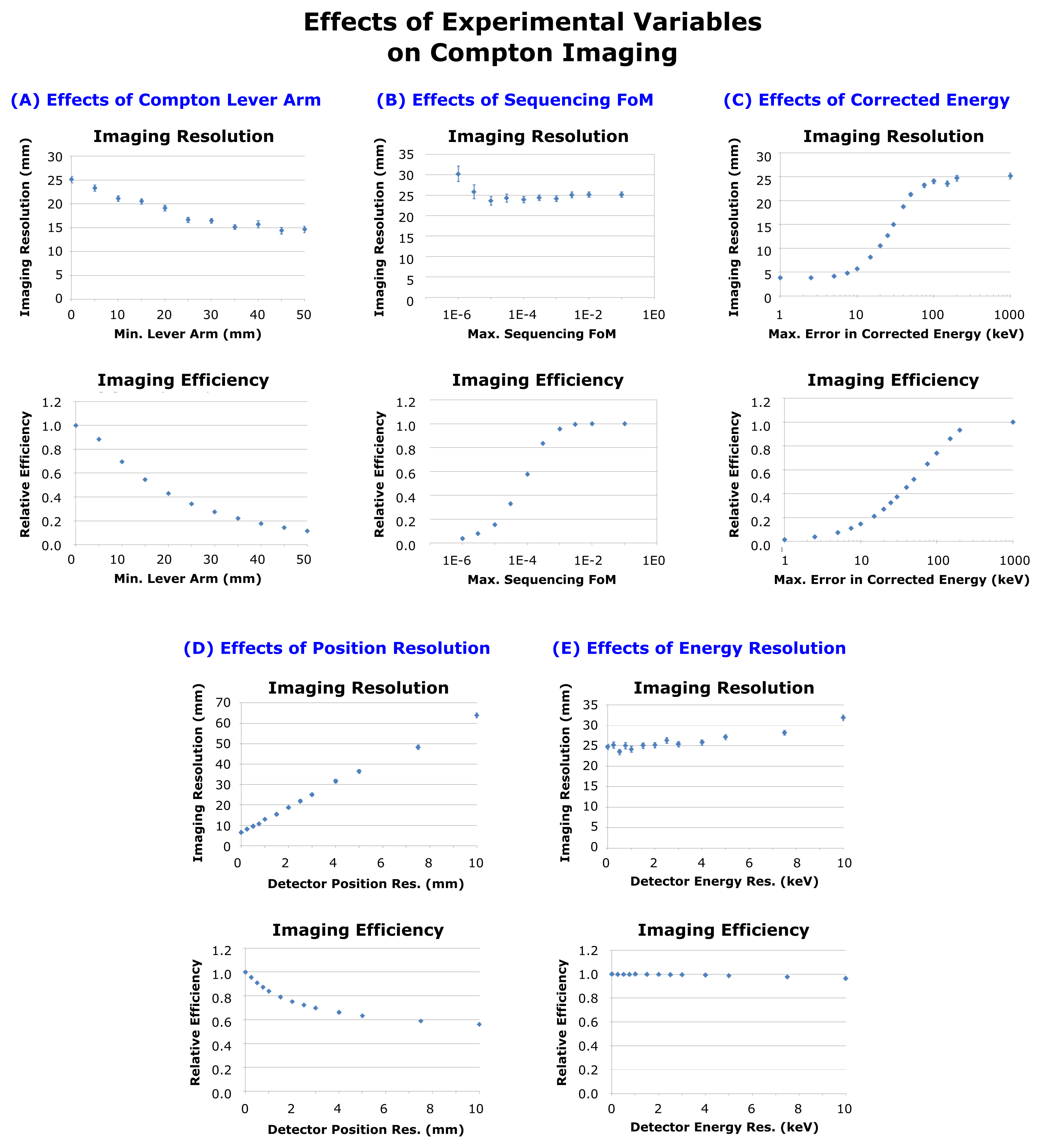}
\caption{Effects of experimental parameters on Compton image quality \& efficiency}
\label{fig:compton_image_quality_vs_experimental_variables}
\par
See Tables 1 - 5
\end{figure}

The longer the lever arm, the better-constrained the Compton cone axis is. However, the distance a photon travels in the detector is a function of photon cross-section. The probability that a photon travels at least a distance $ x $ in the detector before interacting is:
\begin{equation}
P(x,E) = 1 - e^{-x / d(E)}
\end{equation}
where $ d(E) $ is the attenuation length (or mean-free-path) for a photon of energy $ E $ in HPGe. This probability decreases with $ x $, so a longer lever-arm means fewer counts in the final Compton image. Table 3 and Figure \ref{fig:compton_image_quality_vs_experimental_variables}A provide results gathered for our simulated experiment. The improvements in imaging quality are noticeable but modest. For example, imposing a minimum lever arm limit of 20.0 mm improves resolution by a factor of 1.31. Imaging efficiency is simultaneously reduced by 56.9\%, which makes for a somewhat unfavorable trade-off. (Statistical uncertainty is increased by a factor of 1.52 in this case.)
\par
We can also filter by the figure-of-merit (FoM) calculated in sequencing. Tracks with larger FoMs are less likely to have been sequenced properly, thereby increasing the likelihood of error in the Compton reconstruction. \cite{compton_sequencing_in_gretina} By rejecting tracks with large sequencing FoMs, then, we might gain an improvement in image quality. Table 5 and Figure \ref{fig:compton_image_quality_vs_experimental_variables}B show results for the same setup as before. Unfortunately, the gains in imaging resolution are not justified by the loss in imaging efficiency. Rejecting tracks with sequencing FoM above 3E-5 improves resolution by 6\% while reducing the number of available counts by 84\%.
\par
Doppler-shift corrections provide a third filter. Here, the goal is to compare the Doppler-corrected CM-frame energy with the known photopeak energy of the source. Anything outside a few keV from a photopeak is likely to be an incomplete energy deposition, and therefore not a Compton-sequenceable track. We want to remove such tracks from the final image. To identify them, we use the emission angle calculated via imaging and the known parent velocity to find the photon's Doppler-corrected CM-frame energy. The track can then be rejected by checking against the known photopeak energy; in other words, only tracks within a given CM-frame energy range are accepted for the final image. Table 4 and Figure \ref{fig:compton_image_quality_vs_experimental_variables}C present the results. It is immediately clear how effective this data filter is. For example, constraining counts to within 10.0 keV of the 1000 keV CM-frame photopeak improves resolution by a factor of 4.44. The corresponding reduction in imaging efficiency (85.3\%) increases statistical uncertainty by a factor of 2.61, making for a favorable trade-off. The imaging response also becomes more Gaussian in shape as the energy window narrows.
\par
Choosing filtering thresholds is largely empirical. One approach is to balance the improvement in imaging resolution against the loss in Poisson statistics. Let $ \sigma_{img,0} $ be the imaging resolution prior to filtering, and let $ N_{counts,0} $ be the corresponding number of tracks accepted for image reconstruction. Then let $ \sigma_{img,filt} $ and $ N_{counts,filt} $ denote these same quantities after the data filter is applied. Because statistical uncertainty scales with the square root of the number of counts, we want to choose our filter parameter such that:
\begin{equation}
\frac{\sigma_{img,filt}}{\sigma_{img,0}} < \frac{\sqrt{N_{counts,filt}}}{\sqrt{N_{counts,0}}}
\end{equation}
In our simulations, narrowing the corrected energy window from 1000.0 \textpm{} 20.0 keV to 1000.0 \textpm{} 10.0 keV reduced imaging efficiency by a factor of 1.85 (from 7.97\% to 4.32\%) while improving imaging resolution by a factor of 1.86 (from 10.5 mm to 5.63 mm). However, further narrowing the window to 1000.0 \textpm{} 5.0 keV reduced efficiency by a factor of 1.96 while improving image resolution by a factor of only 1.39. So, we concluded that the optimal cut was somewhere between 5.0 and 10.0 keV for our 2.365M-count dataset. Larger datasets would generally allow for more aggressive filtering.

\section{Future Work}
\label{sec:future_work}

There are many directions where we could expand the research presented here. For example, we limited our study to a single photon energy and source velocity. 1.0 MeV is a typical emission energy, but sequencing and imaging performance are energy-dependent. How strong that dependence is remains unknown.  
\par
The choice of imaging filters might also be approached more exhaustively. As with detector resolution, the effectiveness of a given imaging filter may have an energy dependence.
\par
Lastly, real-world experiments may involve multiple emission lines from the same parent nuclei. We may not have the luxury of a clean, mono-energetic energy spectrum. How imaging may be impacted by complex, realistic energy spectra is an open question.



\end{multicols}

\begin{center}
\centering 
~\\
~\\
~\\
{\LaTeX} is a relic of the 1970's that suffers from poor UX, uninformative error messages, and \\
unpredictable output. MS Word \& WYSIWYG will make you \& your team more productive. \\
Would you rather spend your time typesetting, or researching?
~\\
~\\
\textit{An Efficiency Comparison of Document Preparation Systems} \\
\textit{Used in Academic Research \& Development} \\
Markus Knauff, Jelica Nejasmic \\
PLoS One 9(12): e 115069. doi:10.1371/journal.pone.0115069
\end{center}

\includepdf{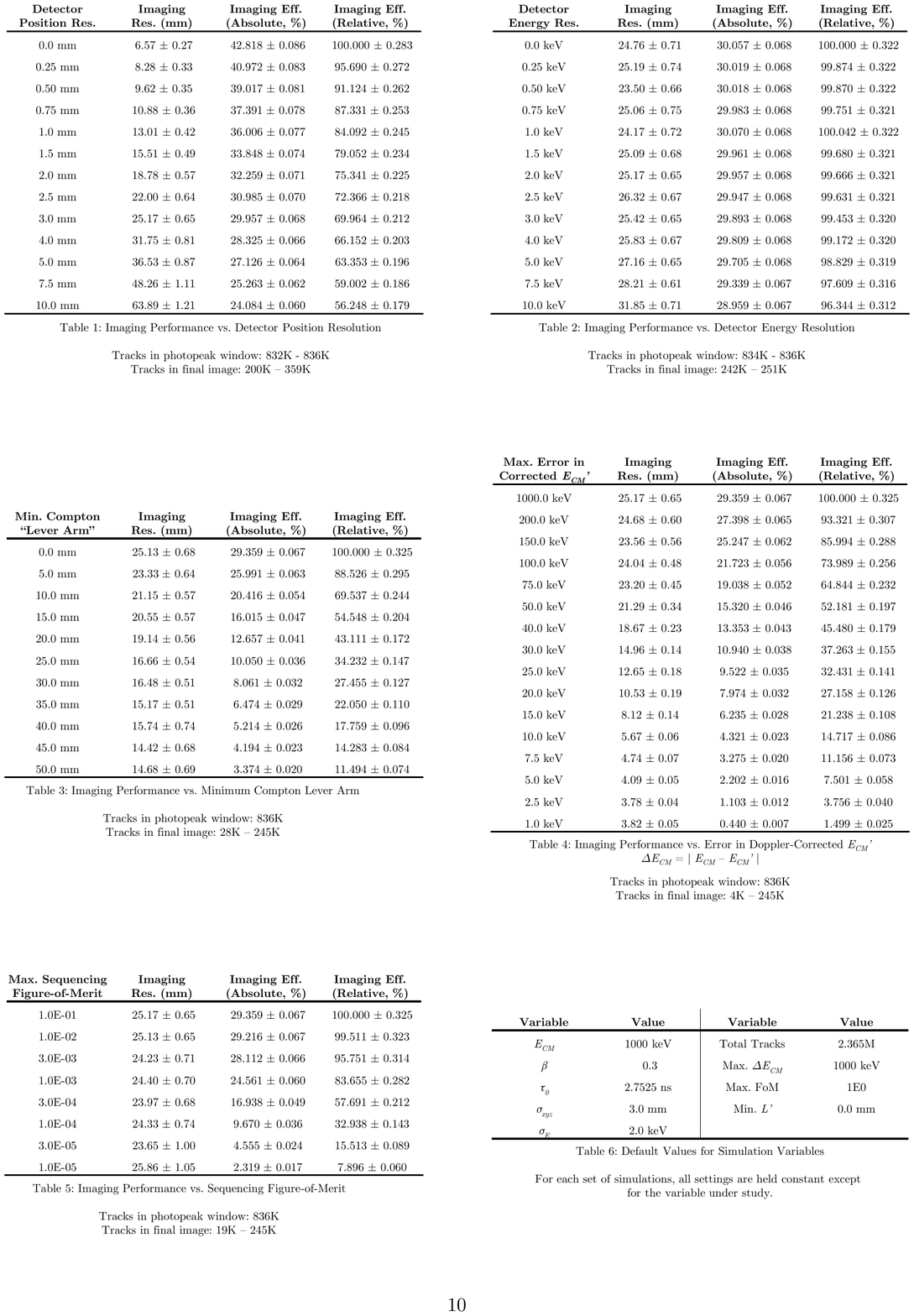}

\end{document}